# THz phonon spectroscopy of doped superconducting cuprates

Ya. G. Ponomarev, Hoang Hoai Van, S. A. Kuzmichev, S.V. Kulbachinskii,
M. G. Mikheev, M. V. Sudakova, S. N. Tchesnokov

Department of Low Temperature Physics, Moscow State University, 119991 Moscow, Russia

Abstract

THz phonon spectroscopy studies of bismuth cuprates are overviewed. Facts are presented evidencing a strong electron–phonon interaction in doped BSCCO superconductors. A pronounced fine structure in dI/dV- characteristics of Josephson break junctions in Bi-2212 and Bi-2223 has been observed which is caused by a nonlineare interaction of AC Josephson current with Raman-active optical phonon modes. For Bi-2223 the range of frequencies extended up to 25 THz, including the apex oxygen mode $O_{Sr}$ with the energy close to 80 meV and two-phonon resonances $(O_{Sr} + Sr)$ and $(O_{Sr} + Cu)$ with the energies 96 meV and 104 meV respectively. At the same time no traces of the magnetic resonance were observed, which allows to conclude that the magnetic resonance doesn't participate in the formation of superconducting properties of bismuth cuprates. The observed "quantization" of the "gap" voltage $V_{gn} = n(2\Delta/e)$ for natural nanosteps on the cryogenically cleaved surfaces of Bi-2212 and Bi-2223 ($\mathbf{j} \parallel \mathbf{c}$) proves the existence of the intrinsic Josephson effect ($\mathbf{n}$ is the number of contacts in a nanostep). A sharp extra structure in the current-voltage characteristics (CVC's) of nanosteps is attributed to the presence of the extended van Hove singularity (EVHS) close to the Fermi level in slightly overdoped and slightly underdoped samples. A superconducting gap in bismuth cuprates is found to scale with $T_c$ on doping. An observed giant instability in I(V) – characteristics of Bi-2212 and Bi-2223 ($\mathbf{j} \parallel \mathbf{c}$) nanosteps is probably caused by a resonant emission of $2\Delta$ – optical phonons in a process of recombination of nonequilibrium quasiparticles (Krasnov - Schnyder model).

## INTRODUCTION

According to Abrikosov **[1]** a high transition temperature $T_c$ in layered cuprates is caused by the presence of an extended van Hove singularity (EVHS) near the Fermi level. In Abrikosov's model, optical phonons with small wave vectors play a dominant role in pairing. The electron-phonon coupling in high-temperature superconductors (HTSC) was confirmed experimentally via the pump-probe optical spectroscopy [ 2 ], the effect of renormalization of the quasiparticle density of states [3-8], the effect of generation of optical phonons by AC Josephson current [9-12], the photoemission spectroscopy [13. 14] and the isotope effect [15].

It is well known that the transition temperature $T_c$ for cuprates varies with the hole concentration, p. according to a parabolic law: $T_c = T_{c,max}[1 - 82.6(p - 0.16)^2]$ (the doping level p is defined as the hole density per Cu site normalized to one $CuO_2$ plane) [16, 17]. At the same time there is a contradictory information about a doping dependence of a superconducting gap $\Delta(p)$.

In the present investigation a transition from overdoped (OD) to underdoped UD) samples was achieved by substituting Sr with La [18]. The intrinsic Josephson effect in nanosteps on cryogenically cleaved surfaces of doped Bi-2212(La) single crystals, Bi-2212 whiskers and Bi-2223 polycrystals has been studied. A discreet character of the "gap" voltage $V_{gn} = n(2\Delta/e)$ for natural nanosteps ($j \parallel c$) was observed ($n$ is the number of contacts in a nanostep). The obtained results support a previously observed scaling of the superconducting gap $\Delta$ with the critical temperature $T_c$ on doping [19]. In addition, we have registered a sharp extra structure in the current-voltage characteristics (CVC's) of perfect Bi-2212 nanosteps which could be caused by the presence of the EVHS [20-23] close to the Fermi level in both slightly overdoped and slightly underdoped samples. An interaction between AC Josephson current and Raman-active optical phonon modes in the entire range of phonon frequencies (up to 25 THz) was observed in doped Bi-2212 and Bi-2223 samples . pointing to a strong electron–phonon coupling in HTSC [24-26]. A giant instability in I(V) – characteristics of Bi-2223 nanosteps is explained by a resonant emission of $2\Delta$ – optical phonons in a process of recombination of nonequilibrium quasiparticles (Krasnov – Schnyder model [27, 28]).

## EXPERIMENTAL

A break-junction technique [29] has been used to generate contacts of SIS-type (tunnelling spectroscopy) and SnS-type (Andreev spectroscopy) in Bi-2212(La) single crystals, Bi-2212 whyskers and Bi-2223 polycrystals. The magnitude of the gap measured by both methods on the same sample usually coincided within experimental errors (Fig.1). In addition, using the same technique we have studied the current – voltage characteristics (CVC's) of natural nanosteps with a height from 1.5 to 30 nm which are always present on cryogenically cleaved surfaces of bismuth cuprates. In most cases these nanosteps were shunted by a single SIS contact, which allowed to estimate the number of contacts $n$ in a stack (Fig. 2 – dI/dV-characteristic of a Bi-2212(La) nanostep (8 contacts) at T = 4.2K (**a**) shunted by a single SIS contact (**b**)).

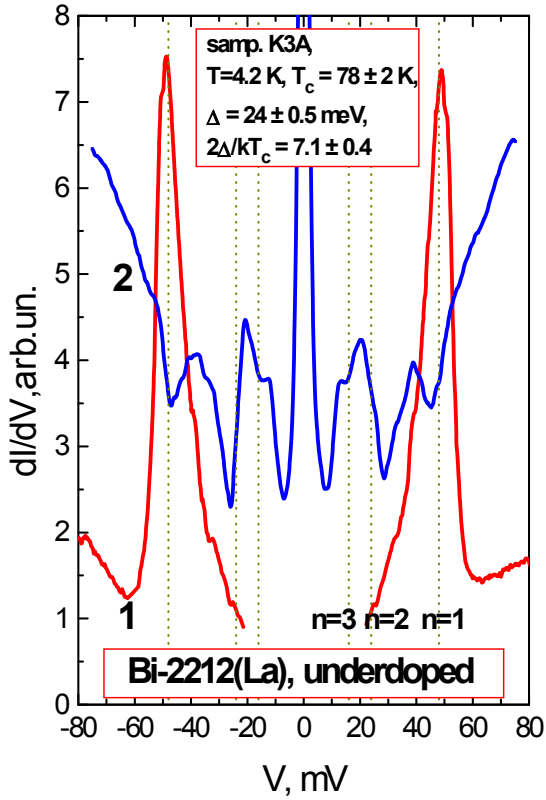

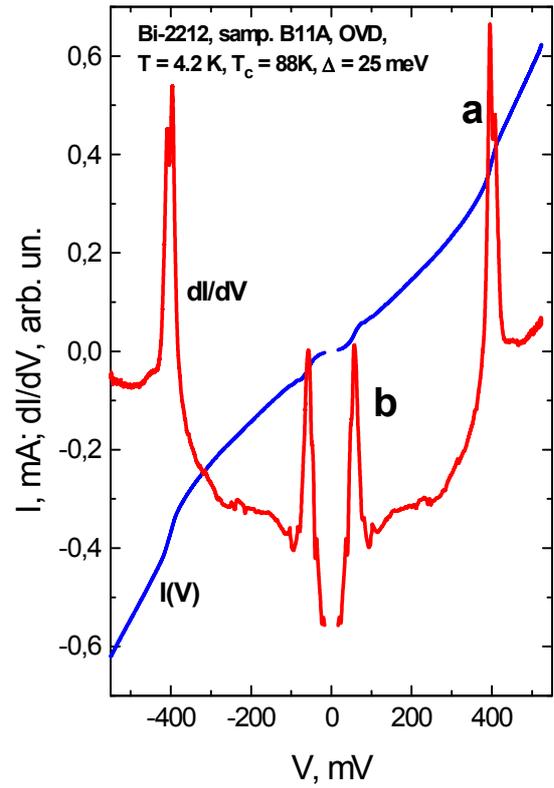

**Fig. 1.** dI/dV- characteristics of a sinle tunneling SIS–contact (1) and a sinle Andreev SnS-contact (2) in underdoped Bi-2212(La) at T = 4.2K ($T_c$ = 78±2 K, $\Delta$ = 24±0.5 meV, $2\Delta/kT_c$ = 7.1±0.4).

**Fig. 2.** dI/dV- characteristics of a Bi-2212(La) nanostep (8 contacts) at T = 4.2K (**a**) shunted by a single SIS contact (**b**) (slightly overdoped sample, $T_c$ = 88K, $\Delta$ = 25.0±0.5 meV).

We have used the data of tunneling, intrinsic tunneling and Andreev spectroscopies to derive the dependence of a superconducting gap $\Delta$ on the impurity hole concentration **p** (Fig. 1 - Fig. 4). It should be noted, that for underdoped samples of bismuth cupratrs it becomes progressively difficult to prepare **true** single SIS contacts with the current in c-direction. The layered structure of the material and weakening of bonding between superconducting blocks causes the formation of complicated network of contacts in the vicinity of the surface. In the present investigation the electron-phonon resonances were used as reliable calibration marks, which helped to judge a single SIS contact from a stack Dotted lines in Fig. 3 indicate the position of electron-phonon resonances $V_{res}$, corresponding to a nonlinear interaction of AC

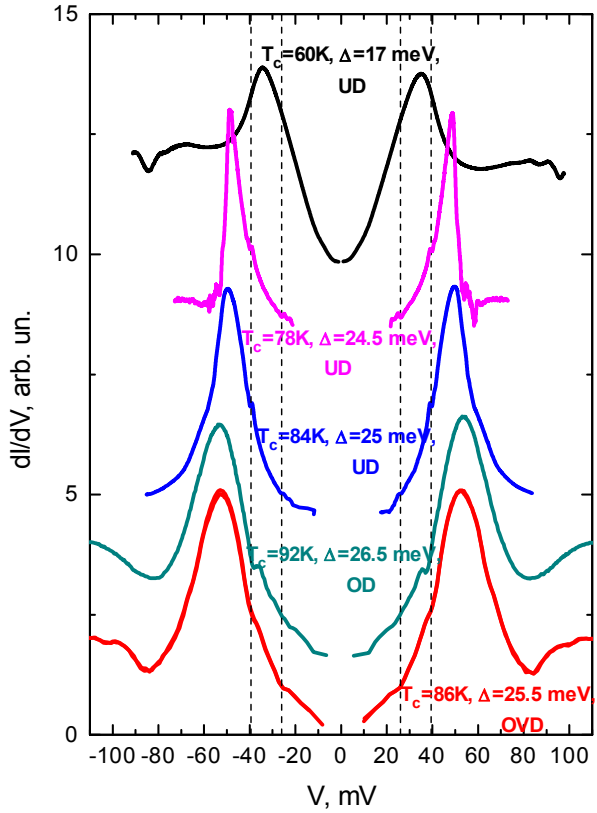

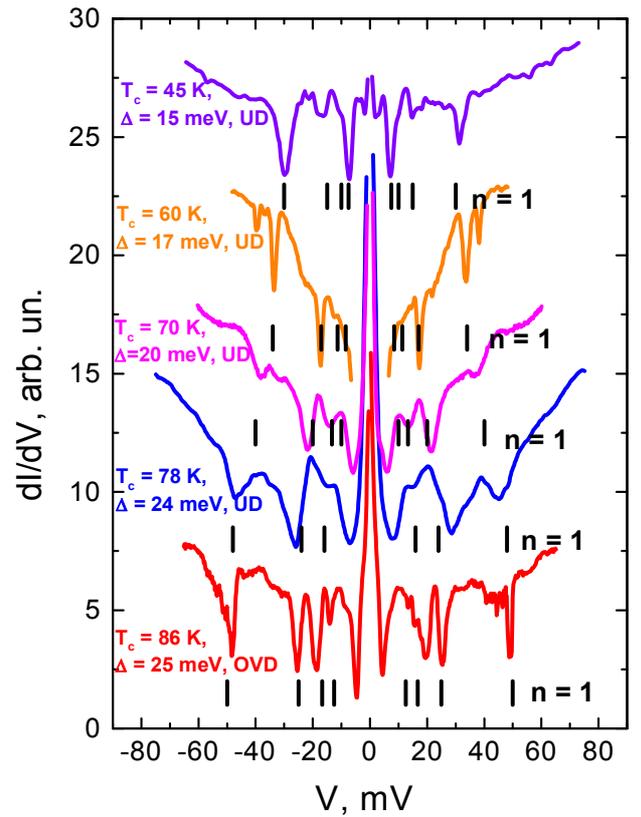

**Fig. 3.** dI/dV–curves for SIS contacts in underdoped Bi-2212.

**Fig. 4.** Resistive transitions R(T) and gaps $\Delta$(T) in underdoped Bi-2212 samples.

Josephson current with Raman- active optical phonon modes [30]. For an apical oxygen phonon mode $O_{Sr}$ : $E_{phon} = 2eV_{res} \approx 80$ meV [9].

In all cases the superconducting gap $\Delta$(T = 4.2 K) was found to scale with the transition temperature $T_c$ (Fig. 5 and Fig. 6) in the entire doping range (the ratio $2/\Delta kT_c$ is close to 7). The flattening of the "gap" parabola in the vicinity of optimal doping (Fig. 6) is probably an indication of pinning of the Fermi level to the EVHS [31].

We have found, that La substitution (red solid circles in Fig. 6) does not affect the character of scaling of $\Delta$ and $T_c$ (empty squares correspond to as grown crystals without La). A similar dependence was reported in [19], though a conflicting experimental information can also be found [32]. Recently scaling of superconducting gap $\Delta$ and $T_c$ was also observed in [33].

The results obtained jn the present investigation and in [19, 33] support the conclusion that the frequency range of HTSC Josephson contacts (important for THz applications) is maximal for optimal doping .

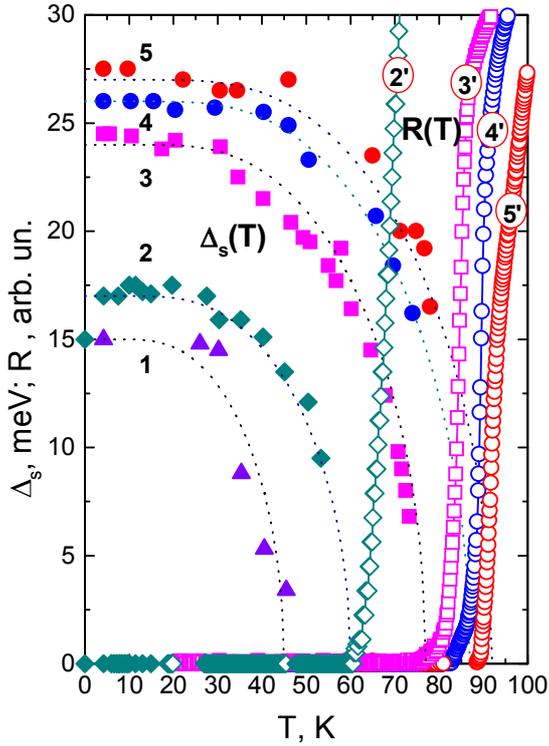

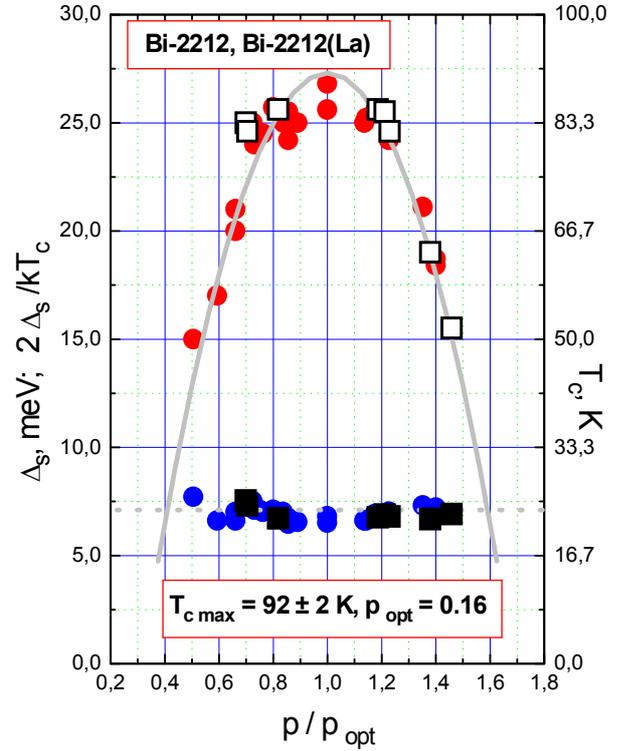

**Fig. 5.** Resistive transitions R(T) and gaps Δ(T) in underdoped Bi-2212 samples.

**Fig. 6.** Doping dependence of Δ and $T_c$ for Bi-2212.

Direct measurements done by the scanning tunneling microscopy (STM) method showed that the height of nanosteps on cryogenically cleaved surfaces is proportional to half the height of the unit cell: 1,5 nm (the cleavage surface is located between two neighbouring BiO planes) [18, 19]. Note that half a unit cell in the c-direction corresponds to a single Josephson junction. According to Kaneko et al. [34] and Mitchell et al. [35], the nanostep width does not exceed 1 μm. This result coincides with the estimates made in Ref. [19]. By tuning the junction with a micrometric screw in a single experiment it was possible to move from one nanostep to another and record their CVC's individually.

The gap voltage for nanosteps with different number of SIS contacts **n** is "quantized" : $V_{gn} = n(2\Delta/e)$ (Fig. 7, Fig. 8), which proves the applicability of the "intrinsic Josephson effect" scenario for cuprates [36]. The sharpness of the gap structure is typical for Bi-2212(La) and Bi-2223 nanosteps and allows to estimate $V_{gn}$ with sufficient accuracy. No signs of overheating

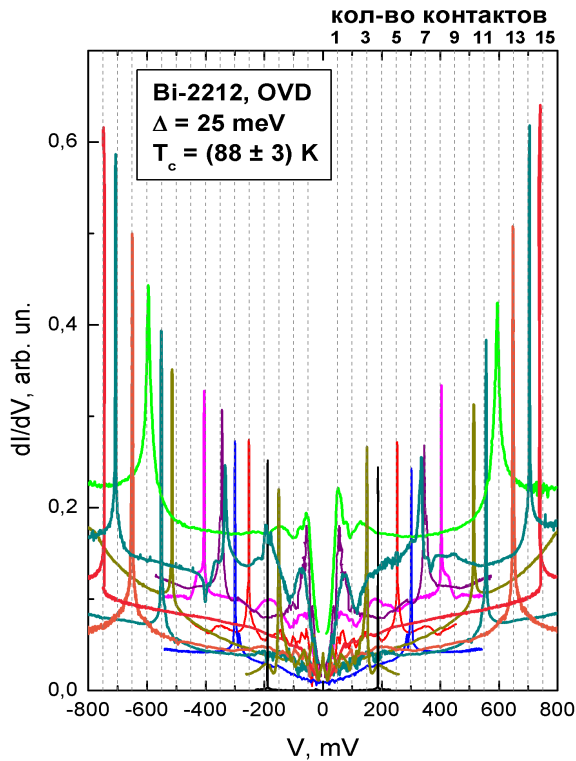 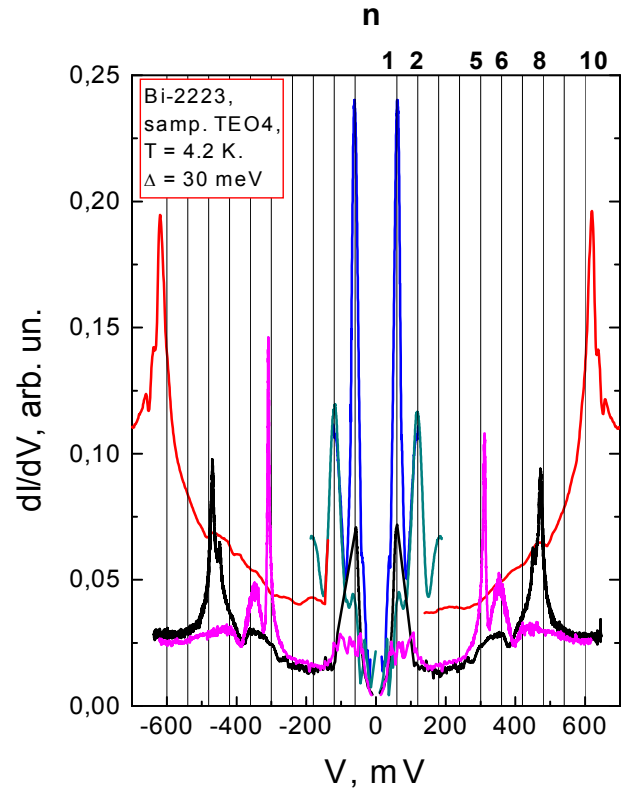

**Fig. 7.** dI/dV–curves for nanosteps with different number of contacts **n** at T = 4.2 K (Bi-2212., $T_c$ = 88 ± 3 K, Δ = 25 meV). The discreteness of the gap voltage $V_{gn}$ = n·(2Δ/e) is clearly seen.

**Рис. 8.** dI/dV–curves for nanosteps with different number of contacts **n at** T = 4.2 K (Bi-2223., $T_c$ = 104 K, Δ = 30 meV). The discreteness of the gap voltage $V_{gn}$ = n·(2Δ/e) is clearly seen.

at biase voltages V ≈ $V_{gn}$ were observed. For slightly underdoped Bi-2212(La) single crystal in Fig. 7 an elementary gap voltage 2Δ/e equals 50 mV at T = 4.2 K..

The gap feature in the CVC's has a shape typical for an 's-symmetry' (isotropic) gap parameter. At the first glance it is difficult to match this result with the photoemission spectroscopy data, according to which the gap parameter in the ab plane is highly anisotropic [37]. However , the situation changes when there is a van Hove singularity at the Fermi level. Wei et al. [38] had found that a van Hove singularity enhances the gap structure in the CVC's of junctions even when the gap parameter in the ab plane is highly anisotropic.For slightly overdoped and slightly underdoped samples a sharp extra structure in the CVC's of Bi-2212 nanosteps was observed (Fig.9, Fig. 10), which we attribute to the presence of the extended van

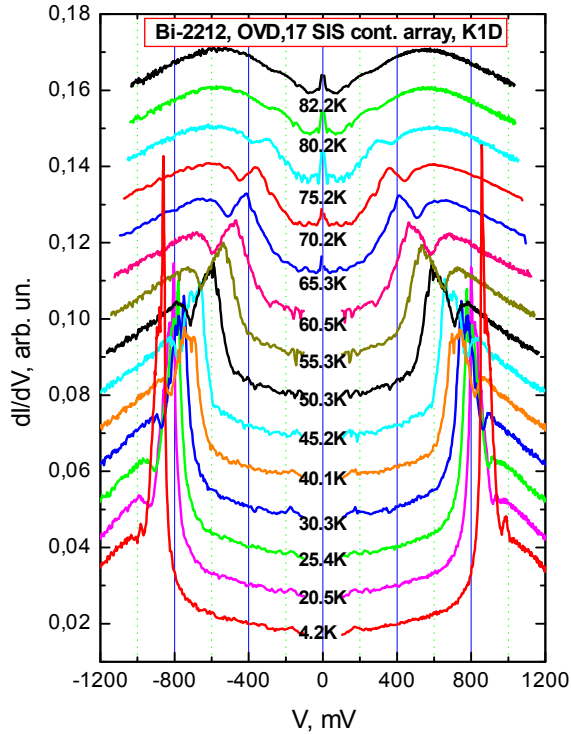

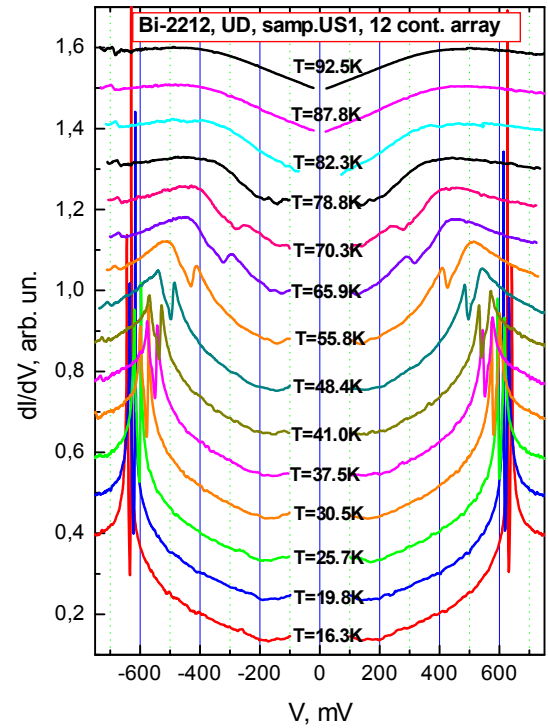

**Fig. 9.** dI/dV-characteristics of an overdoped Bi-2212 nanostep with 17. contacts at different temperatures.

**Fig. 10.** dI/dV-characteristics of an underdoped Bi-2212 nanostep with 12 contacts at different temperatures.

Hove singularity (EVHS) close to the Fermi level [1, 20, 21, 23, 39]. This extra structure is absent in the CVCs of optimally doped samples (Fig. 2) where it falls into a $2\Delta$ – region. For strongly underdoped and strongly overdoped samples the extra structure is also not detectable. The latter effect could be the result of a pronounced drop of a quasiparticle lifetime above the superconducting gap [40].

The evolution of CVC's with temperature in Fig.9 and Fig. 10 agrees qualitatively with theoretical calculations of Bok and Bouvier [21] (van Hove scenario). The inner singularity in Fig.9 and Fig. 10 (and in [21]) closes at $T_c$ and obviously corresponds to the gap structure, the outer singularity does not disappear at $T_c$ and may be caused by EVHS (see also a recent publication: A. Piriou et al., First direct observation of the Van Hove singularity in the tunneling spectra of cuprates, Nature Communications 2, 221, 01 March 2011 [23]).

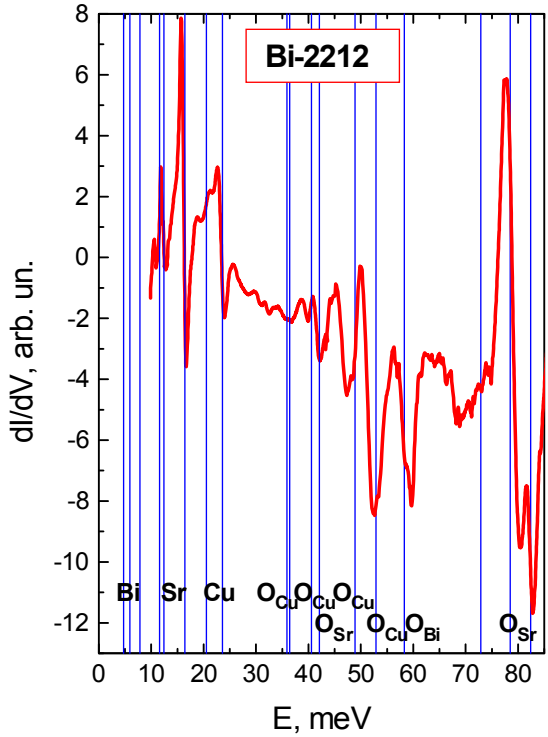

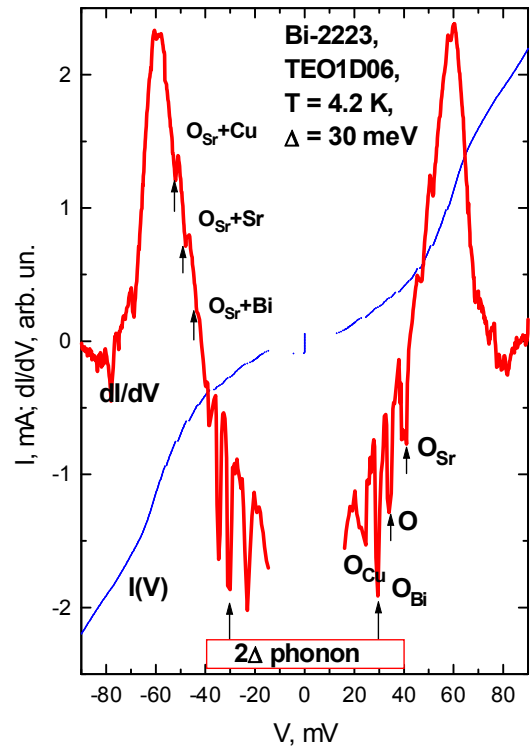

**Fig.11.** Phonon resonances on a dI/dV-characteristic of a break junction in optimally doped Bi-2212 at T = 4.2 K (E = 2e·V). Vertical <u>blue</u> lines correspond to the energies of Raman-active phonon modes [9].

**Fig.12.** I(V) – and dI/dV – characteristics of a break junction in Bi-2223 (T = 4.2 K, $\Delta$ = 30 meV). Single-phonon and two-phonon resonances are marked by arrows.

Fig. 11 presents a dynamic conductance dI/dV of a break-junction with a partially suppressed background as a function of energy E = 2e·V (optimally doped Bi-2212 single crystal, T = 4.2K). A pronounced structure in the dI/dV – characteristic (Fig. 11) is caused by an interaction of AC Josephson current with Raman-active optical phonon modes [30] in the range of phonon frequencies up to 20 THz, including the apex oxygen mode with the energy close to 80 meV. Vertical dashed lines in Fig. 11 correspond to the energies of Raman-active optical phonon modes (from Ref. [9]).

It has been found that the structure in dI/dV-characteristics of Bi-2212 Josephson junctions related to generation of optical phonon modes $\omega_i$ at bias voltages V = $\hbar\omega_i$/2e can be observed for both underdoped and overdoped Bi-2212 and Bi-2223 single crystals (Fig. 11 and Fig. 12), with the doping level having only a negligible effect on the frequency of the main phonon modes (Fig. 13). This means that the electron-phonon coupling in BSCCO does not change significantly with doping over the entire region where superconductivity exists. Our experimental data support a model of a strong electron–phonon interaction [1, 24-26] in HTSC. At the same time we have not observed any structure which could be related to the "magnetic resonance" mode whose energy is expected to change significantly on doping [41, 42] (green rectangles in Fig. 13 mark expected positions of a "magnetic resonance" [42] for every sample). Thus the role of magnon modes in pairing is questionable.

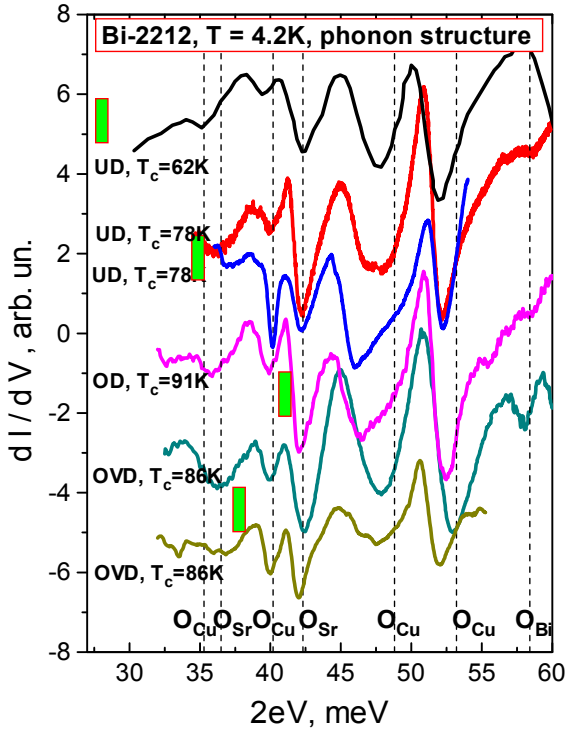

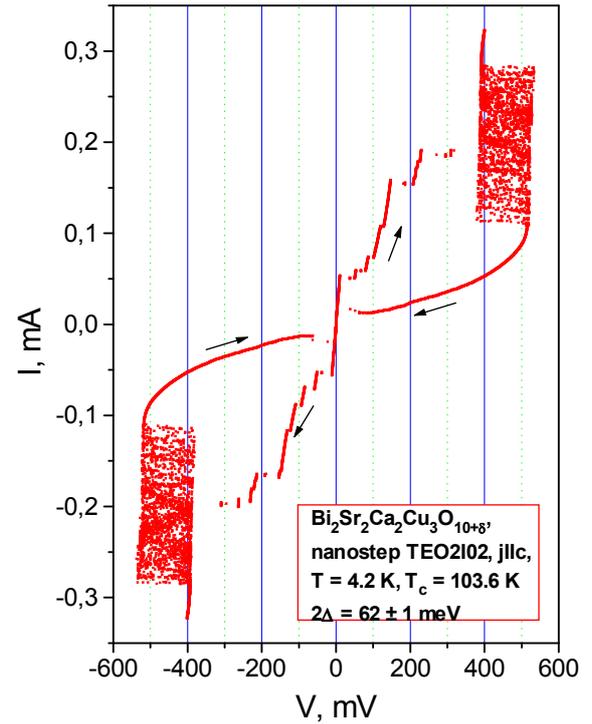

**Fig. 13.** Phonon resonances on dI/dV-characteristics of break junctions in doped Bi-2212 at T = 4.2 K.

**Fig. 14.** I(V) – characteristic of a Bi-2223 nanostep with a gaint instability at T = 4.2 K

A giant instability in I(V) – characteristics of Bi-2223 nanosteps has been observed at helium temperatures (Fig.14). A periodic switching between 7th and 9th branches in a multibranch characteristic presented in Fig. 14 could be caused by a resonant emission of 2Δ – optical phonons with the energy ≈ 60 meV ( $O_{Bi}$ – mode in Fig. 11 and Fig. 12) in a process of recombination of nonequilibrium quasiparticles in a stack of SIS junctions (Krasnov - Schnyder model [27, 28]). Two main factors were found to influence the amplitude of the instability of this kind. The instability amplitude reaches its maximum when: 1) EVHS crosses the Fermi level, 2) the value 2Δ coincides with the energy of one of the optical phonon modes. The instability amplitude is very sensitive to temperature and quickly drops in a "step-like" way when equality $2\Delta(T) = \hbar\omega_{opt.phon.}$ is not valid any more. For Bi-2223 phase the $2\Delta$ –value at helium temperature does not exceed 70 meV (optimal doping) and the "apical oxygen" mode ($\sim$ 80 meV) remains beyond reach. At the same time, using chemical doping we can tune the $2\Delta$ – value at least to two optical phonon modes with energies $E_{phon} \cong 60$ meV ($O_{Bi}$ mode in BiO – planes, Fig. 11 and Fig. 12) and $E_{phon} \cong 50$ meV ($O_{Cu}$ mode in $CuO_2$ –planes). We succeeded to observe instabilities in CVCs of B-2223 nanosteps for both cases. For Bi-2212 phase the $2\Delta$ – value does not exceed $52 \div 54$ meV (optimal doping) and instability easily develops in Bi-2212 nanosteps close to optimal doping.

## 3. Conclusions

Tunneling, intrincic tunnelling and Andreev spectroscopy studies of doped Bi-2212 and Bi-2223 samples are overviewed. Facts are presented evidencing the importance of the electron–phonon interaction in HTSC. A sharp extra structure in the CVCs of Bi-2212 contacts is attributed to the presence of the extended van Hove singularity close to the Fermi level in slightly overdoped and slightly underdoped samples. A giant instability in I(V) – characteristics of Bi-2223 and Bi-2212 nanosteps could be probably an indication of phonon pairing in HTSC .
.

**Acknowledgements** The authors are grateful to V. M. Pudalov and L. M. Fisher for useful discussions. This work was supported by the Russian Foundation for Basic Research (project nos. $11 - 02 - 01201$, 08-02-00935 and 05-02-17868).

Boockholt M., Buschmann L., Gunherodt G., Physica C 235- 240 (1994) 1863.